\documentclass[twocolumn,10pt,amssymb,prl,aps,superscriptaddress,longnamesfirst]{revtex4-1}

\usepackage{lipsum}
\usepackage{graphicx,amsmath}
\usepackage{epstopdf}
\usepackage{xcolor}
\usepackage[normalem]{ulem}
\usepackage{pgfplots}
\usetikzlibrary{positioning}
\usetikzlibrary{decorations.text}
\usetikzlibrary{decorations.pathmorphing}
\usetikzlibrary{arrows,arrows.meta,shapes.geometric}
\usepgfplotslibrary{groupplots}
\usepgfplotslibrary{fillbetween}
\pgfplotsset{compat=newest}
\pgfplotsset{plot coordinates/math parser=false}

\newcommand{\expect}[1]{\langle #1 \rangle}

\newcommand{\figref}[1]{Fig.~\ref{#1}}

\newcommand{\cqcct}{Centre for Quantum Computation and Communication Technology}
\newcommand{\engaddr}{Research School of Engineering, The Australian National University, Canberra, ACT 2600, Australia}
\newcommand{\dqsaddr}{Department of Quantum Science, Research School of Physics and Engineering, The Australian National University, Canberra, ACT 2600, Australia}
\newcommand{\UQaddr}{School of Mathematics and Physics, University of Queensland, St. Lucia, Queensland 4072, Australia}

\begin{document}
 \title{Violation of Bell’s inequality using continuous variable measurements}
 \author{Oliver~Thearle}
 \affiliation{\cqcct}
 \affiliation{\dqsaddr}
 \affiliation{\engaddr}
 \author{Jiri~Janousek}
 \affiliation{\cqcct}
 \affiliation{\engaddr}
 \author{Seiji~Armstrong}
 \author{Sara~Hosseini}
 \affiliation{\cqcct}
 \affiliation{\dqsaddr}
 \author{Melanie~Sch\"unemann~(Mraz)}
 \affiliation{Arbeitsgruppe Experimentelle Quantenoptik, Institut f{\"u}r Physik, Universit{\"a}t Rostock, D-18051 Rostock, Germany}
 \author{Syed~Assad}
 \affiliation{\cqcct}
 \affiliation{\dqsaddr}
 \author{Thomas~Symul}
 \author{Matthew~R.~James}
 \author{Elanor~Huntington}
 \affiliation{\cqcct}
 \affiliation{\engaddr}
 \author{Timothy~C.~Ralph}
 \affiliation{\cqcct}
 \affiliation{\UQaddr}
 \author{Ping~Koy~Lam}
 \affiliation{\cqcct}
 \affiliation{\dqsaddr}

\begin{abstract}
A Bell inequality is a fundamental test to rule out local hidden variable model descriptions of correlations between two physically separated systems. There have been a number of experiments in which a Bell inequality has been violated using discrete-variable systems. We demonstrate a violation of Bell’s inequality using continuous variable quadrature measurements. By creating a four-mode entangled state with homodyne detection, we recorded a clear violation with a Bell value of $B=2.31\pm0.02$. This opens new possibilities for using continuous variable states for device independent quantum protocols.
\end{abstract}
\maketitle
A Bell test is a fundamental demonstration of quantum mechanics. It is made up of a family of inequalities that test the hypothesis of local realism \cite{Brunner2014}. Violation of a Bell inequality between spatially separated sub-systems demonstrates that there exist non local correlations between them. Only entangled quantum systems can violate a Bell inequality in this way. This has application in quantum technologies where one can be faced with the verification of quantum devices. For quantum key distribution (QKD) and quantum random number generators (QRNG) a violation of a loop-hole free Bell inequality can rule out a compromise of the quantum source or measurement devices by a third party. This allows the users to achieve device independent (DI) protocols \cite{Pironio2009}. 

In quantum optics there are two ways to decompose the optical field. One is to quantize the optical field into discrete photon numbers. This allows information to be encoded in discrete variables (DV). These systems can have very low bandwidths from photon generation and high detection losses at room temperature \cite{Lita2008} but they are relatively robust to channel losses and noise. Bell inequalities have been violated with DV systems for over 35 years \cite{Aspect1982} with ever increasing efficiency. These violations have relied on the ``fair-sampling" assumption - a loop-hole that could be exploited by an adversary. With the recent improvement in photon detection efficiencies at cryogenic temperatures there have been three significant demonstrations of a loophole free Bell test \cite{Giustina2015,Hensen2015,Shalm2015}. These experiments will allow
for true DV DI-QRNG \cite{Pironio2010} and DI-QKD \cite{acin2007} protocols. 

The second approach, used in this letter, is to consider a decomposition into the continuous variable (CV) amplitude and phase quadratures of the optical field. The advantages of CV systems are that high detection efficiency is much easier to achieve and the resource states are deterministically generated. For CV quantum optics a Bell test is harder to realize. Bell argued that Quantum states with positive-definite Wigner functions would not violate a Bell inequality with respect to CV measurements \cite{Bell2004}. This seems to rule out the use of commonly produced two-mode CV entangled states. These states are widely known as EPR states. There have been several protocols proposed which try to use more exotic states with photon subtraction \cite{Garcia2004} or using photon-wave correlations \cite{Valerio2011}. However, it was shown in Ref.~\cite{Ralph2000} that in fact it is possible to violate a Bell inequality with EPR states using CV measurements provided one trusts the measurement system. In this letter we experimentally demonstrate such a Bell state violation.
\begin{figure}[]
	\centering
	\includegraphics{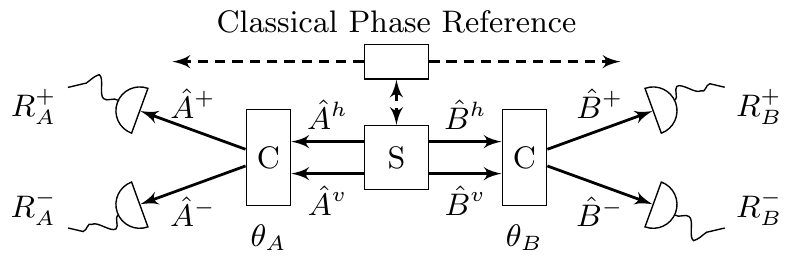}
	\caption{A diagram of a generic Bell test.}
\label{fig:generic}
\end{figure}
Typical protocols use some variation of the classic Bell test protocol depicted in Fig.~\ref{fig:generic}. The source, S, generates a four mode correlated optical state; two parties Alice and Bob are then given two modes each, $\hat{A}^h$, $\hat{A}^v$ and $\hat{B}^h$, $\hat{B}^v$ separated in polarization. They can mix their two modes using the mixer, C, to perform one of two measurements $\{\theta_A,\theta'_A\}$ and $\{\theta_B,\theta'_B\}$ on their modes. Measuring the resulting modes $\hat{A}^+$, $\hat{A}^-$, $\hat{B}^+$ and $\hat{B}^-$ with single photon detectors will give one of two outcomes, $R\in\{0,1\}$. Repeating this experiment a number of times Alice and Bob can build up correlation statistics between each others measurement outcomes with,
\begin{equation}
	R^{ij}(\theta_A,\theta_B) =\expect{R_A^i(\theta_A)R_B^j(\theta_B)},
\label{eq:photoncorr}
\end{equation}
where $i,j\in\{+,-\}$. The expectation value of the correlation for each of the four combination of measurement settings is given by,
\begin{widetext}
\begin{equation}
E(\theta_A,\theta_B) = \frac{R^{++}(\theta_A,\theta_B) + R^{--}(\theta_A,\theta_B) - R^{+-}(\theta_A,\theta_B) - R^{-+}(\theta_A,\theta_B)}{R^{++}(\theta_A,\theta_B) + R^{--}(\theta_A,\theta_B) + R^{+-}(\theta_A,\theta_B) + R^{-+}(\theta_A,\theta_B)}.
\label{eq:E}
\end{equation}
\end{widetext}
These expectations can then be used to form the well known CHSH inequality \cite{Bell2004},
\begin{equation}
B = |E(\theta_\mathrm{A},\theta_\mathrm{B}) + E(\theta'_\mathrm{A},\theta'_\mathrm{B}) + E(\theta'_\mathrm{A},\theta_\mathrm{B})
 - E(\theta_\mathrm{A},\theta'_\mathrm{B})| \leq 2.
\label{eq:B}
\end{equation}
This inequality places a bound on what is possible with local realism models and can only be violated using entangled states. A maximal violation of the inequality can be observed with measurement settings $\theta_A = \{\frac{\pi}{8},\frac{3\pi}{8}\}$ and $\theta_B = \{0,\frac{\pi}{4}\}$. This basic protocol can also be varied to have more parties, measurements or outcomes \cite{Brunner2014}.

The continuous variable Bell test proposals in Ref.~\cite{Ralph2000,Huntington2001} are based around an entanglement source using optical parametric oscillators (OPO) and homodyne measurements. The photon number correlations needed for a Bell test are inferred through the homodyne quadrature measurements using the equivalence,
\begin{equation}
\hat{A}^\dagger\hat{A} \equiv (\hat{A}^\dagger \hat{A} - \hat{V}^\dagger\hat{V}) = \frac{1}{4}(\hat{X}_A^2+\hat{P}_A^2 - \hat{X}_V^2 - \hat{P}_V^2),
\label{eq:equiv}
\end{equation}
for a mode $\hat{A}$. Here the quadrature operators are defined as $\hat{X}_F=\hat{F}+\hat{F}^\dagger$ and $\hat{P}_F=i(\hat{F}^\dagger - \hat{F})$ in terms of the annihilation and creation operators $\hat{F}$ and $\hat{F}^\dagger$ for mode $F\in\{A,V\}$ where $V$ is the background vacuum mode with the corresponding creation operator $\hat{V}^\dagger$. The measurement of the background vacuum is inherent in homodyne measurement and a direct measurement of the detected field will yield Eq.~\eqref{eq:equiv}. If Alice and Bob consider the photon number in each detected mode the correlation equation Eq.~\eqref{eq:photoncorr} becomes, 
\begin{equation}
R^{ij} =\expect{\hat{A}^\dagger_i\hat{A}_i\hat{B}^\dagger_j\hat{B}_j}.
\end{equation}
Using the equivalence relation Eq.~\eqref{eq:equiv}, the correlation Eq.~\eqref{eq:photoncorr} can be rewritten again to be in terms of homodyne quadrature measurements. By assuming Gaussian statistics, all correlations can be reduced to second order correlations. In this case, using $\expect{\hat{X}^2 \hat{Y}^2}=\expect{\hat{X}^2} \expect{\hat{Y}^2} +2 \expect{\hat{X} \hat{Y} }^2$, we have
\begin{align}
R^{ij} = &\frac{1}{16}[2(\expect{\hat{X}^i_{A}\hat{X}^j_{B}}^2 + \expect{\hat{P}^i_{A}\hat{P}^j_{B}}^2 + \expect{\hat{X}^i_{A}\hat{P}^j_{B}}^2 + \expect{\hat{P}^i_{A}\hat{X}^i_{B}}^2)\nonumber\\ &+ V^i_{A;X}V^j_{B;X} + V^i_{A;P}V^j_{B;P} + V^i_{A;P}V^j_{B;X}+ V^i_{A;X}V^j_{B;P} \nonumber\\
&- 2V_v(V^i_{A;X} + V^i_{A;P}) - 2V_v(V^j_{B;X} + V^j_{B;P})\nonumber\\
&+ 4V_v^2].
\label{eq:corrfun}
\end{align}
Here $V^i_{F;X}=\expect{(\hat{O}_F)^2}$ for $\hat{O}\in\{\hat{X},\hat{P}\}$ where $F$ is the mode $A$, $B$ and $V_v$ is the second moment of the vacuum mode. To see how Eq.~\eqref{eq:corrfun} can be used to produce a Bell violation the significance of each term can be explored \cite{Ralph2000,Huntington2001}. The first four terms are dependent on the measurement angle with the next four being polarization independent. The last three terms come from the quantum noise of the vacuum state. In a perfect experiment the polarization independent terms will cancel with the quantum noise terms to create a high correlation fringe visibility with respect to $\theta_\mathrm{A}$ and $\theta_\mathrm{B}$. This fringe visibility can be diminished by the measurement of uncorrelated photons from classical noise sources and high order photon number terms such as those in highly entangled CV states. In a purely classical experiment the last three terms will be zero and result in a small correlation fringe.

In regards to this protocol it is assumed that the contribution of the vacuum mode will be such that $\expect{\hat{V}^\dagger\hat{V}}=0$ to meet the requirement that Eq.~\eqref{eq:equiv} remains a positive operator. If this assumption is violated it opens loopholes that could explain a Bell violation from this protocol. To rule out this loophole the photon number count for the $\hat V$ mode, i.e. with all the light blocked, $n_\mathrm{dark}$, should be much less than the photon number count in the local oscillator, $n_\mathrm{LO}$. In particular $n_\mathrm{dark}\ll\sqrt{n_\mathrm{LO}}$. This test demonstrates that the homodyne measurements are truly of vacuum correlations. It is well established by many experiments that this is a good assumption at optical frequencies. However this requires trust of the detection device.

\begin{figure}[t]
	\centering
	\includegraphics[width=\columnwidth]{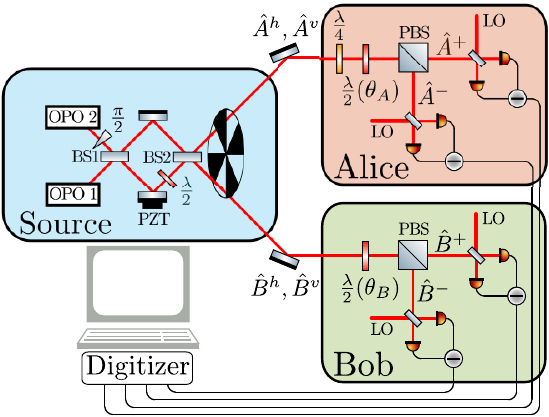}
\caption{A schematic diagram of the experiment. The Bell state is generated by mixing two orthogonal squeezed states in the same linear polarization on a 50:50 BS to generate an EPR state. One arm of the EPR state is rotated into the orthogonal polarization. The two beams are then interfered on BS2 to give four correlated modes; $\hat{A}^h$, $\hat{A}^v$, $\hat{B}^h$ and $\hat{B}^v$ separated spatially and in polarization. Alice and Bob each receive two polarization separated modes and mix their received modes by $\theta_A$ and $\theta_B$ respectively. The resulting modes; $\hat{A}^+$, $\hat{A}^-$, $\hat{B}^+$ and $\hat{B}^-$ are measured with homodyne detectors.}
\label{fig:expsch}
\end{figure}

To observe a violation of Eq.~\eqref{eq:B} with the correlation function Eq.~\eqref{eq:corrfun} a CV source is required to produce an entangled state. For this experiment we have used the third source proposed in Ref.~\cite{Huntington2001} based on the well known Bell test performed by Ou and Mandel \cite{Ou1988}. Rather than post-selecting entangled photons by photon counting as in Ref.~\cite{Ou1988}, we analyse the CV correlations of a similar state according to Eq.~\eqref{eq:corrfun}. As shown in Fig.~\ref{fig:expsch} the entangled state was created by interfering two orthogonal squeezed states on a 50:50 BS (BS1). The squeezed states were created in the side bands of spatially separated beams of a Nd:YAG 1064nm laser. The side bands were squeezed using two singly resonant bow tie cavity OPO's each containing a 1cm long periodically poled Potassium Titanyl Phosphate (ppKTP) crystal. Both of the OPO's were seeded by the 1064nm laser. A second harmonic generator provided 532nm pump for the ppKTP crystals to create the squeezed light. The Bell state was created by changing the entangled states into orthogonal linear polarization to be then mixed on a second 50:50 BS (BS2). The four modes are then distributed with $\hat{A}^h$ and $\hat{A}^v$ to Alice and $\hat{B}^h$ and $\hat{B}^v$ to Bob. Alice and Bob then respectively mixed their polarization separated states by angles $\theta_A$ and $\theta_B$ using a half wave plate ($\lambda/2$) and a polarizing beam splitter (PBS). The resulting states, $\hat{A}^+$, $\hat{A}^-$, $\hat{B}^+$ and $\hat{B}^-$ were then measured using homodyne detectors.

As this experiment is derived from a discrete variable Bell test the result will be invariant to relative phase between each beam path. However it is necessary to lock each homodyne to orthogonal quadratures. To do this the experiment used two phase modulations applied separately to the OPO's for Pound Drever Hall locking. The phase between each of the beam paths was controlled by a piezo controlled mirror to hold the modulations orthogonal to each other. An additional quarter wave plate ($\lambda/4$) was used to correct for a phase miss match between $\hat{A}^h$ and $\hat{A}^v$ caused by BS2. 

The correct measurement of shot noise in this experiment is crucial to ensure relationships underpinning this Bell test. In particular ensuring Eq.~\eqref{eq:equiv} remains a positive operator. The laser intensity was found to drift up to $1\%$ over the course of the experiment. To measure the correct shot noise an optical beam chopper was used to rapidly switch the homodyne detectors between measuring the signal and shot noise. This reduced the requirement on the stability of the experimental setup. Incorrectly measuring shot noise can lead to spurious violations of Eq.~\eqref{eq:B} for unentangled states.
\begin{figure}[t]
	\centering
	\includegraphics{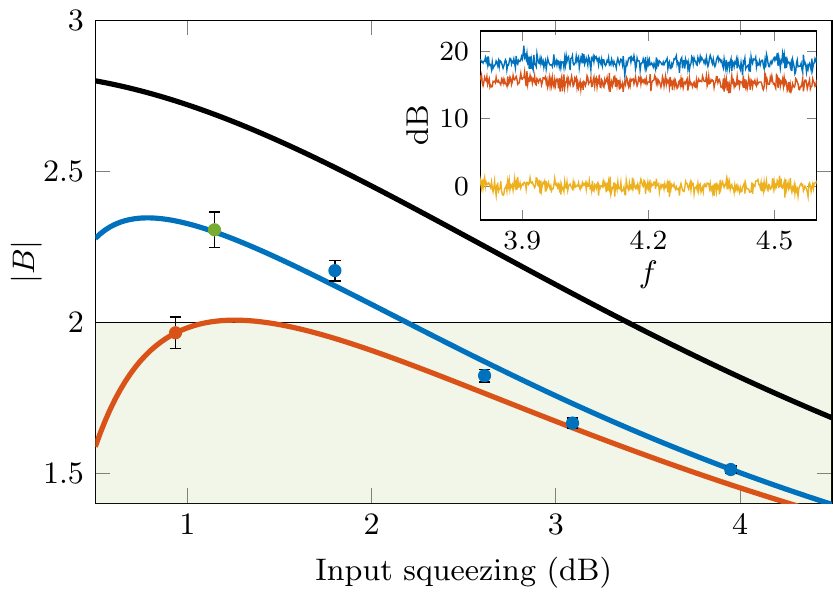}
	\caption{Bell violations showing the effect of different experimental parameters vs the inferred value of input squeezing. The largest violation was $B = 2.31$ (green point) at 15 standard deviations above the classical limit (shaded) with a measured detector dark noise of 17.5 dB using 5.2 mW of LO power. Increasing the input squeezing decreases the violation (blue points). Decreasing the dark noise clearance by decreasing the LO power to 2.6 mW pushes $|B|$ below $2$ (red point). The error bars shown are three standard deviations of the mean. The solid lines are of the fitted model described by Eq.~\eqref{eq:modeleq} for each of the dark noise clearances. The theoretical maximum violation is given by the black line. The LO power spectral density estimate relative to the dark noise (yellow) is shown in the inset for 5.2 mW (blue) and 2.6 mW (red) of LO power for the sideband of interest.
\label{fig:B}}
\end{figure}
\begin{figure*}[t]
\centering
	\includegraphics{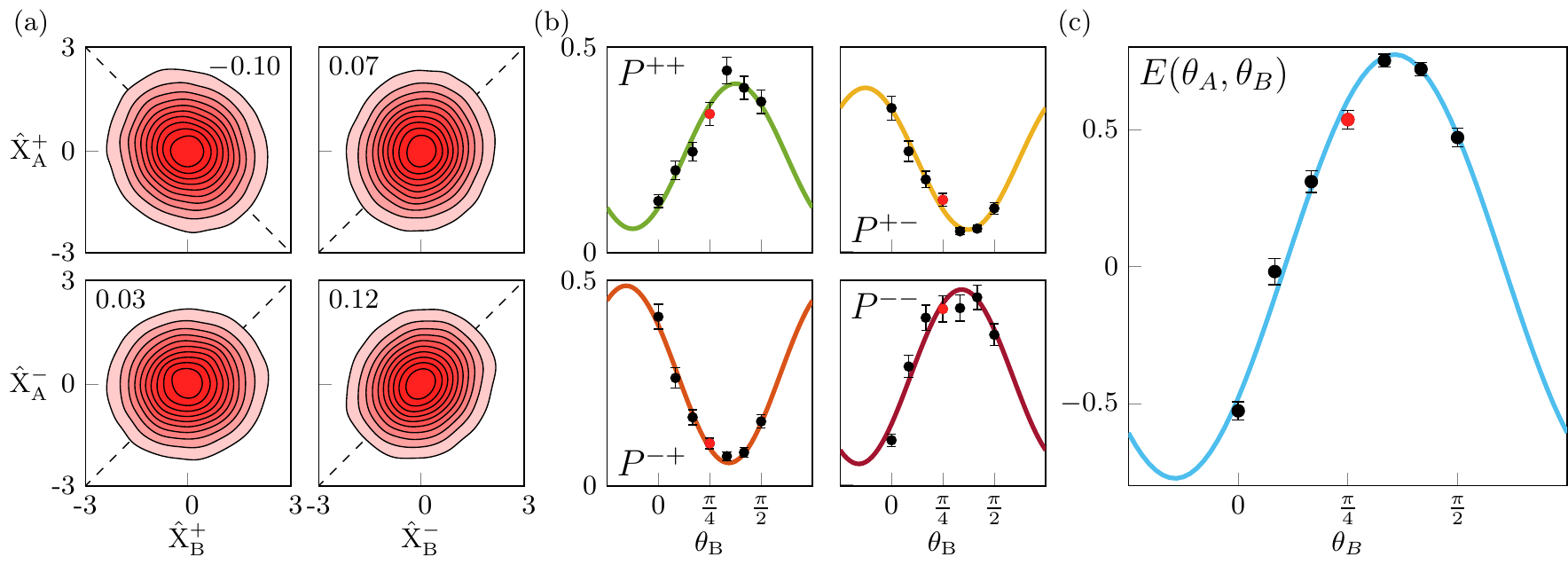}
	\caption{Raw homodyne data correlations collected with 1.1 dB of input squeezing represented in a topographical map with weak Pearson correlations, (a), translates to strong photon correlation (red points in (b)). The photon correlations fringes in (b) are found with Eq.~\eqref{eq:pequation} with the recorded visibility above $75\%$ when $\theta_\mathrm{B}$ is swept with $\theta_\mathrm{A}=\pi/8$. The correlations are then used to find the expectation fringe (c) using Eq.~\eqref{eq:E}. The solid lines are fitted using the model described by Eq.~\ref{eq:modeleq}.
\label{fig:visplot}}
\end{figure*}

Modeling of the experiment with this source shows several important factors that could reduce the Bell violation \cite{Huntington2001,Ralph2000}. Underlying this Bell test is essentially a single photon experiment and as such the inequality will be maximally violated when the source mostly produces correlated pairs of single photons. An important parameter is then the input squeezing; with high levels of input squeezing the Bell violation decreases. A squeezed state decomposed into the Fock basis reveals that the state is made from photons in sets of multiples of two with a decreasing probability. Increasing squeezing of a state will increase the probability of the higher order photon terms occurring. These terms can introduce correlations that dilute the Bell correlations and decrease the violation. Any noise in the experiment will have the same effect of decreasing the violation though by decreasing the correlations. The two main sources of noise for this experiment were identified as the input state purity and detector dark noise. 

In the single photon equivalent, experiment loss will only increase the number of samples required to get a significant correlation value. However, for this experiment the loss will also decrease the violations by increasing the effect of noise that appears at the output such as detector dark noise. From modeling it was found the maximal violation for the experimental setup used would occur with both OPO's generating approximately $1$ dB of squeezing with a dark noise clearance of 17.5~dB below shot noise measured for each homodyne detector with $\approx5.2$ mW of local oscillator power.

A set of four fixed measurement settings were identified that would give all the correlation and variance terms required by Eq.~\eqref{eq:corrfun}. These measurements were made in a fixed order for each combination of $\theta_\mathrm{A}$, $\theta'_\mathrm{A}$, $\theta_\mathrm{B}$ and $\theta'_\mathrm{B}$ with the shot noise regularly sampled during measurements. The dark noise measurement was only taken once at the end of each experimental run. As we were not seeking to address loopholes the detectors were located next to each other and sampled using the same digitizer. 

The main result presented in this letter is the violation of Eq.~\eqref{eq:B} with $B=2.31$ with a standard deviation of 0.02 with $1.1$ dB of inferred input squeezing found by bootstrapping the data. The violation of Eq.~\eqref{eq:B} was also demonstrated with up to $1.8$ dB squeezing of the input field. Sweeping of the input squeezing with both OPO's in \figref{fig:B} shows the effect of increasing the anti-squeezing noise on the experimental setup. As the OPO's are pumped harder to produce more squeezing the purity of the state they produce decreases due to more noise in the anti-squeezed quadrature. This purity decreased from $0.98$ for $1.1$~dB of squeezing to $0.92$ for $3.9$~dB of squeezing. The results from Ref.~\cite{Huntington2001} show that for a similar detector noise it should be possible to observe a Bell violation for up to $3$~dB of squeezing, a result not observed in this experiment due to the decreasing purity of the squeezed states. A second experimental run was conducted where the local oscillator power for each detector was decreased from 5.2 mW to 2.6 mW to simulate the effect of an increase in detector dark noise. This gave the expected result of a decrease in violation of Eq.~\eqref{eq:B}. The values of squeezing quoted here were inferred from fitting the model described by Eq.~\ref{eq:modeleq} and agree well with the direct measurements of the input squeezing. Each Bell violating was measured on a sideband centered at 4.2 Mhz with a bandwidth of 1 Mhz.

A third experimental run was conducted to observe the correlation fringe. To do this $\theta_A$ was fixed at $\pi/8$ while $\theta_B$ was swept from $0$ to $\pi/2$ rad. The input squeezing was set to be 1.1 dB. The correlation fringes from this experiment are plotted in \figref{fig:visplot} (b) as normalized $P$ values. The $P$ values are calculated with,
\begin{equation}
P^{ij}=\frac{R^{ij}}{\sum_{i,j} R^{ij}}.
\label{eq:pequation}
\end{equation}
The correlation fringe visibility was measured to be over $75\%$. This could be further improved by reducing the noise in the experiment. From the normalized P values we can draw a comparison with the recorded homodyne data plotted in \figref{fig:visplot} (a) with the corresponding Pearson correlation. For the raw homodyne data a very weak correlation is observed but from this a significant P value is still observed. The process of calculating $B$ is given a visual representation by reading \figref{fig:visplot} from left to right. The homodyne correlations and variances are used to calculate the photon correlations and then the expectation value for each measurement setting.

For \figref{fig:B} and \figref{fig:visplot} (b) and (c) a model was fitted to the experimental data. The Gaussian assumption made for Eq.~\ref{eq:corrfun} meant that the experiment could be completely described by an 8 by 8 co-variance matrix, $\gamma$. This matrix was constructed such that each sub matrix $\gamma^{ij}$, where $i,j\in\{2n-1,2n\}$, represents the two quadratures for one of the four measured modes indexed by $n$. Using $\gamma_\mathrm{in}$ to represent the input state each element in this experiment is applied using a sympletic operation with the matrix operation $\gamma=S\gamma_\mathrm{in} S^T$. To add the contribution of efficiency, $\eta$, and noise relative to the output, $\varepsilon$, a completely positive map \cite{Raul2008} was used to arrive at,
\begin{equation}
	\gamma = \sqrt{\eta}\mathbb{I} S  \gamma_{in} S^T \sqrt{\eta}\mathbb{I} + \varepsilon \mathbb{I}.
	\label{eq:modeleq}
\end{equation}
The input state, $\gamma_{in}$, was taken to be the state after the OPO's with its diagonal given by the vector $\begin{matrix}[V_{sqz} & V_{asqz} & V_{sqz} & V_{asqz}&1&1&1&1]\end{matrix}$. Here $V_{sqz}$ is the variance of the squeezed quadrature and $V_{asqz}$ the anti-squeezed quadrature. This model was fitted to each of the experimentally obtained values of $R^{ij}$ using an iterative fitting process to find $\eta$, $\varepsilon$, $V_{asqz}$ and $V_{sqz}$. The measured parameters provided the starting point for the fitting with the input squeezing measured directly on a homodyne by using mirrors to bypass the optical nextwork in Fig.~\ref{fig:expsch}. 

In this letter we have demonstrated the first observation of Bell correlations in a continuous variable system with a violation of 2.31 at 15 standard deviations above the classical limit with a detector dark noise of 17.5dB below shot noise. This result demonstrates the strength of photon number correlations when inferred through homodyne measurements. A demonstration of a violation of the Bell inequality was also made with 1.8 dB of input squeezing and would be possible to up to 2 dB of input squeezing with this experiment. These correlations exist between side-band modes of a bright beam that would be very difficult to measure directly via photon counting. This result was possible because of the high correlation fringes observed with this experiment. While this Bell test fails to address any loopholes it is still a significant result as a proof of principle for CV Bell tests. In order for this violation to be believed the detection devices must be trusted due to the hard to close loop-hole caused by the shot-noise verification. Never-the-less this Bell test could be applied to a source independent QRNG similar to those protocols proposed in Ref.~\cite{ma2016,Marangon2017}.

\begin{acknowledgments}
We would like to thank Valerio Scarani for his helpful discussions. This research is supported by the Australian Research Council (ARC) under the Centre of Excellence for Quantum Computation and Communication Technology (Project No. CE110001027).
\end{acknowledgments}

\bibliography{bibfile}

\end{document}